\documentclass[conference]{IEEEtran}
\IEEEoverridecommandlockouts
\usepackage{cite}
\usepackage{amsmath,amssymb,amsfonts}
\usepackage{algorithmic}
\usepackage{graphicx}
\usepackage{textcomp}
\usepackage{xcolor}
\def\BibTeX{{\rm B\kern-.05em{\sc i\kern-.025em b}\kern-.08em
    T\kern-.1667em\lower.7ex\hbox{E}\kern-.125emX}}
    
\begin{document}

\title{Advanced Topic Modeling Techniques for Categorizing Software Vulnerabilities}

\author{
\IEEEauthorblockN{
Utkarsh Tiwari\textsuperscript{1}, Spoorthi M\textsuperscript{2}, Anirudh S\textsuperscript{3}, and Nidhin Prabhakar T. V.\textsuperscript{4*}}
\IEEEauthorblockA{
Department of Computer Science \& Engineering, \\
Amrita School of Computing, Bengaluru, \\
Amrita Vishwa Vidyapeetham, India \\
\textsuperscript{1}bl.en.u4cse21212@bl.students.amrita.edu, 
\textsuperscript{2}bl.en.u4cse21193@bl.students.amrita.edu, \\
\textsuperscript{3}bl.en.u4cse21020@bl.students.amrita.edu, 
\textsuperscript{4*}tv\_nidhin@blr.amrita.edu
}
}

\maketitle

\begin{abstract}
The increasing complexity and frequency of software vulnerabilities demand efficient methods to analyze and prioritize threats. Traditional approaches often fail to process the vast amount of unstructured textual data effectively, highlighting the need for advanced solutions. This study leverages state-of-the-art topic modeling techniques powered by large language models (LLMs) to extract meaningful insights from the ’Threat’ feature of a software vulnerability dataset. Models such as BERTopic, Top2Vec, CombinedTM, Llama2 with BERTopic, and Mixtral are utilized, along with dimensionality reduction and clustering methods like UMAP, PCA, HDBSCAN, and DBSCAN. By uncovering latent patterns and generating interpretable clusters, this research enhances threat prioritization and decision-making in cybersecurity. The findings support scalable and automated solutions for vulnerability management, contributing to improved security practices.
\end{abstract}

\begin{IEEEkeywords}
Software Vulnerabilities, Topic Modeling, Large Language Models, Cybersecurity, BERTopic, Top2Vec, CombinedTM, Llama2 with BERTopic, Mixtral, Dimensionality Reduction, Clustering
\end{IEEEkeywords}

\section{Introduction}

Software vulnerabilities pose a threat to the security of various organizations which may lead to financial, reputational, and operational risks. With time, the volume and complexity of vulnerabilities grow, and it is necessary to identify and categorize them effectively. Identifying and categorizing these vulnerabilities effectively has become a major challenge for organizations. This paper addresses this gap by employing state-of-the-art topic modeling techniques to analyze and categorize vulnerabilities.

The proposed workflow starts off with preprocessing the dataset to ensure data compatibility. Five advanced topic modeling approaches—BERTopic with multiple configurations, CombinedTM, Top2Vec, Llama2 with BERTopic, and mixtral\_8x7b—are applied to extract latent topics. Comparative analysis evaluates the performance of each model using metrics like topic coherence and clustering quality, supported by visualizations to aid interpretability. This paper benefits various cybersecurity researchers, software developers, and organizations by providing actionable insights into recurring vulnerability themes. The contributions of this paper involve:

\begin{itemize}
    \item \textbf{Application of BERTopic with Multiple Configurations:} The paper explores four unique configurations of BERTopic using UMAP, PCA, DBSCAN, and advanced language embeddings, demonstrating the adaptability of the framework in identifying meaningful patterns.
    
    \item \textbf{Incorporation of CombinedTM and Top2Vec:} CombinedTM aligns Bag-of-Words representations with contextual embeddings, while Top2Vec efficiently identifies topics by embedding documents and words into a shared semantic space without iterative optimization.
    
    \item \textbf{Integration of Large Language Models (LLMs):} Models such as Llama2 with BERTopic and mixtral\_8x7b are used for contextual topic labeling and document-specific topic generation, enhancing the interpretability of results.
\end{itemize}

Moving on, the paper is structured in such a way that Section 2 presents the Literature Survey, Section 3 talks about the proposed methodology, Section 4 discusses Result Analysis and Discussion, which is then followed by the conclusion.
\section{Literature Review}
 Traditional topic modeling techniques, such as Latent Dirichlet Allocation, are limited by poor interpretability and attractiveness of the results for the end user. The very recent works, like GPTopic [1], introduced a dynamic and interactive way to enhance the interpretability of topics in an easier manner for users. Similarly, other approaches in which clustering techniques are integrated with large language models, such as in the work of Petukhova et al. [3], demonstrate a great promise for better embedding into text data of meaningful clusters. Some other methods, such as those proposed by Frei et al. [4] and Zeng et al. [12], in the cybersecurity domain, have concentrated on improving threat detection with machine learning to enable much faster and more accurate identification of vulnerabilities. Integration of ChatGPT for explanation and elaboration of generated topics could allow users to better understand and satisfy them much more. These innovations pave the way for more efficient and interpretable topic modeling in particular, and real-time clustering and vulnerability detection approaches may be foreseen shortly with further improvement in the present research field. The motivation for further research is to enhance the current models by incorporating LLM-driven approaches and automating processes such as summarization of literature reviews [9] and topic evaluation [10], which in turn will enhance the robustness and applicability of these techniques across various domains, including cybersecurity and multilingual data processing.

Reuter et al. [1] proposes GPTopic, which represents a dynamic and iterative topic modeling method for finally making topic representations more interpretable. The proposed model adopts a Gaussian process framework together with iterative user feedback in order to incrementally update topics. Accordingly, this approach improves not only the satisfaction but also the interpretability of the identified topics over the traditional unsupervised topic modeling solutions. Rijcken et al. [2] makes topic modeling more comprehensible by incorporating ChatGPT into their process. They apply ChatGPT in order to explain topics coming from traditional models in such a way that it does make more sense for a user. Results showed such a method increased the interestingness of topics but also increased users' comprehension and satisfaction in comparison with a standard output. Petukhova et al.[3] discusses the improvement of text clustering by the use of large language model embeddings and further applying clustering methods like k-means and hierarchical clustering. Regarding future work, they would like to investigate multilingual datasets and real-time clustering. Frei et al. [4] discusses the topic of cybersecurity, focusing on the analysis of vulnerability data from software updates and security advisories. The findings have been more effective in the identification of the vulnerabilities in an optimised way, and to automatically adapt threats and integrate some predictive capabilities in the future. Yu et al. [5] presents a hybrid approach for the detection of vulnerabilities in web services, using static code analysis combined with dynamic testing. They intend to extend this approach towards cloud systems and use machine learning to predict emerging vulnerabilities. Akash et al. [6] proposes a system for the enhancement of short-text topic modeling by a combination of LLM-driven context expansion and prefix-tuning of VAEs. This enhances topic coherence, achieving high performances over baseline models. Future enhancements proposed include the optimization of context expansion to domain specificity and further exploration of adaptive prefix tuning. Schneider et. al. [7]  develops a bag-of-sentences approach that fine-tunes LLMs and yields subtle sentence embeddings, which improves topic coherence and relevance. Future work will be related to optimizing this method for different languages and applying it to multilingual datasets. Mu et al.[8] proposes a system that generates interpretable topics directly, eliminating the traditional topic modeling and fine-tuning LLMs to. In that way, clustering and probabilistic models can be avoided, while more coherent and relevant topics are obtained. They further intend to refine the fine-tuning process and scale up the approach for real-time applications in the future. Gana et al. [9], focuses on the automation of literature reviews using LLMs. Their contribution synthesizes key themes and findings from academic papers in a much timelier and comprehensive way than could be done otherwise. Yang et al. [10] describe a procedure for the evaluation of topic models using LLMs, which involves scaling of the evaluation, mostly matching, if not outperforming, human judgment. This work needs further refinement to handle more complex topics and its integration with automated workflows. Sandilya et al. [11] presents an overview of topic-agnostic conversation generation with LLMs, providing a method to naturally conduct flexible conversations across diverse domains. The results indicate increased versatility in conversational agents. Zeng et al. [12] reviews deep learning techniques based on neural networks and reinforcement learning for software vulnerability detection. These methods outperform the traditional approaches in terms of accuracy and scalability. Future research will focus on hybrid models and real-time detection in dynamic environments. Williams et al. [13] presents the framework that uses historical data on vulnerabilities using machine learning to predict whether a piece of software may have vulnerabilities. They planned to extend this work in future work by embedding this framework into continuous integration systems with real-time data.

\section{Methodology}

This paper employs advanced topic modeling techniques to analyze and categorize software vulnerabilities by extracting the latent topics from the "Threat" feature present in the dataset [14]. The process begins with ensuring the data is compatible with the different modeling techniques used in this paper by performing the necessary preprocessing steps. Multiple topic modeling approaches—BERTopic with multiple configurations, CombinedTM, Top2Vec, Llama2 with BERTopic, and mixtral\_8x7b—are applied to the preprocessed data. Each model adopts a unique methodology tailored to its working, resulting in topic clusters or labeled representations of the data. Lastly, a comparative analysis is performed to assess the models' effectiveness in this use case.

\subsection{Model-Specific Methodologies}

\subsubsection{BERTopic with Multiple Configurations}
BERTopic is a highly flexible framework that integrates dimensionality reduction, clustering, and embedding methods to produce interpretable topics. To test its adaptability, four configurations—UMAP, PCA, DBSCAN, and advanced language embeddings—were applied to analyze the "Threat" feature in the dataset. Each setup combined different approaches to dimensionality reduction and clustering, along with pre-trained embeddings, to identify meaningful patterns in the data. The Preprocessing included:
\begin{itemize}
    \item Removing duplicate entries and null values.
    \item Cleaning special characters and punctuations from the text.
    \item Applying minimal stopword removal to preserve contextual richness.
\end{itemize}

\begin{table}[htbp]
\caption{Configurations and Hyperparameters for BERTopic}
\begin{center}
\renewcommand{\arraystretch}{1.3} 
\scriptsize 
\begin{tabular}{|p{1.5cm}|p{1.45cm}|p{1.2cm}|p{2.8cm}|} 
\hline
\textbf{Configuration} & \textbf{Dimensionality Reduction} & \textbf{Clustering Method} & \textbf{Key Parameters} \\
\hline
BERTopic with UMAP & UMAP & HDBSCAN & 
\begin{tabular}[c]{@{}l@{}} 
$n\_neighbors=30$, \\ $n\_components=5$, \\ $min\_cluster\_size=10$
\end{tabular} \\
\hline
BERTopic with PCA & PCA & HDBSCAN & 
\begin{tabular}[c]{@{}l@{}} 
$n\_components=30$, \\ $min\_cluster\_size=10$
\end{tabular} \\
\hline
BERTopic with DBSCAN & UMAP & DBSCAN & 
\begin{tabular}[c]{@{}l@{}} 
$eps=0.5$, \\ $min\_samples=5$
\end{tabular} \\
\hline
BERTopic with Language Embedding & UMAP & HDBSCAN & 
\begin{tabular}[c]{@{}l@{}} 
$n\_neighbors=15$, \\ $n\_components=5$, \\ $min\_cluster\_size=10$
\end{tabular} \\
\hline
\end{tabular}
\label{tab:bertopic_configurations}
\end{center}
\end{table}

The first configuration used UMAP for dimensionality reduction, preserving both global and local structures in the data. HDBSCAN was then applied to the reduced embeddings to group them into meaningful clusters. The second configuration replaced UMAP with PCA, which reduced the embeddings into 30 components, capturing the maximum variance. HDBSCAN was again used for clustering. In the third configuration, DBSCAN replaced HDBSCAN as the clustering method. This approach focused on identifying dense clusters while treating sparse areas as noise. UMAP remained as the dimensionality reduction technique. The fourth configuration incorporated advanced pre-trained language embeddings, combining SentenceTransformer and StackedEmbeddings (RoBERTa with GloVe) to enhance the contextual understanding of the text. UMAP reduced the dimensionality of these embeddings, and HDBSCAN grouped the reduced data into topics.

For all configurations, BERTopic identified top keywords for each topic using term vectorization (CountVectorizer). Visualizations, including heatmaps, hierarchical clustering, and bar charts, were generated to interpret the topics effectively. These outputs provided a clearer understanding of the "Threat" feature and its underlying patterns.

\subsubsection{CombinedTM}
CombinedTM is a hybrid topic modeling approach that combines Bag-of-Words (BoW) representations with contextual embeddings from Transformer models, enabling it to capture necessary statistical and semantic relationships present in the data [15]. The preprocessing steps for CombinedTM are:
\begin{itemize}
    \item Removed special characters, punctuation, and stopwords.
    \item Converted text to lowercase for uniformity.
    \item Generated contextual embeddings using the all-mpnet-base-v2 model.
    \item Prepared a Bag-of-Words matrix.
\end{itemize}

The CombinedTM utilizes a neural network that aligns BoW and contextual embeddings. The network is trained with an objective function that reduces the alignment error between the two representations:

\[
L = -\sum_{i=1}^{N} \log P(\text{BoW}_i | \text{Embedding}_i)
\]

This alignment ensures the latent topics are consistent across both features, yielding coherent topic-word distributions.

\begin{table}[htbp]
\caption{Hyperparameters of CombinedTM}
\begin{center}
\begin{tabular}{|c|c|}
\hline
\textbf{Parameter} & \textbf{Value} \\
\hline
Number of topics & 20 \\
\hline
Contextual embedding size & 768 \\
\hline
Number of epochs & 10 \\
\hline
\end{tabular}
\label{tab:combinedtm_hyperparameters}
\end{center}
\end{table}
\subsubsection{Top2Vec}

Top2Vec has the ability to identify topics by embedding documents and words into a shared semantic space, followed by clustering to form the topics. Unlike CombinedTM, the model Top2Vec does not require iterative optimization, making it resource- and computationally-efficient. The necessary preprocessing steps for Top2Vec are:

\begin{itemize}
    \item Converted text to lowercase.
    \item Retained the most words with minimal cleaning to maintain contextual information.
\end{itemize}

Top2Vec leverages pre-trained embeddings to generate a shared semantic space for documents and words. UMAP reduces the dimensionality of these embeddings, while HDBSCAN clusters the reduced embeddings into dense topic groups. This approach avoids iterative optimization and trains in a single step, focusing on clustering quality. The hyperparameters for Top2Vec are shown in Table~\ref{tab:top2vec_hyperparameters}:

\begin{table}[htbp]
\caption{Hyperparameters of Top2Vec}
\begin{center}
\begin{tabular}{|c|c|}
\hline
\textbf{Parameter} & \textbf{Value} \\
\hline
Speed & "learn" \\
Dimensionality Reduction (UMAP) & Applied \\
Clustering Method & HDBSCAN \\
\hline
\end{tabular}
\label{tab:top2vec_hyperparameters}
\end{center}
\end{table}

\subsubsection{Llama2 with BERTopic}

Llama2 with BERTopic combines embeddings, clustering, and language-based labeling to increase interpretability. It relies on pre-trained embeddings for clusters and uses Llama2 to label topics based on contextual keywords. The preprocessing steps involve:

\begin{itemize}
    \item Removed special characters and punctuation.
    \item Eliminated stopwords for cleaner text.
    \item Generated embeddings using the BAAI/bge-small-en model.
\end{itemize}

The BERTopic framework then reduces the embeddings' dimensionality using UMAP and clusters those reduced embeddings into topics with HDBSCAN. Each cluster is labeled by Llama2-7B-chat, which generates concise topic labels based on the top keywords in each cluster. A predefined prompt template ensures the labels generated are consistent and interpretable, as shown below:

\begin{quote}
[INST]
I have a topic that contains the following documents:
{DOCUMENTS}

The topic is described by the following keywords: '{KEYWORDS}'.

Based on the information about the topic above, please create a short label of this topic. Make sure you only return the label and nothing more.
[/INST]
\end{quote}

The hyperparameters used for this model are highlighted in Table~\ref{tab:llama2_hyperparameters}:

\begin{table}[htbp]
\caption{Hyperparameters of Llama2 with BERTopic}
\begin{center}
\begin{tabular}{|c|c|}
\hline
\textbf{Parameter} & \textbf{Value} \\
\hline
UMAP $n\_neighbors$ & 15 \\
UMAP $n\_components$ & 5 \\
UMAP Metric & "cosine" \\
HDBSCAN $min\_cluster\_size$ & 150 \\
\hline
\end{tabular}
\label{tab:llama2_hyperparameters}
\end{center}
\end{table}

\subsubsection{mixtral\_8x7b}

The mixtral\_8x7b model is a large language model (LLM) used for generating document-specific topics interactively. Unlike traditional models, it does not require clustering or iterative training [16]. It relies on the LLM's pre-trained ability to understand and synthesize natural language through a prompt-driven process. The preprocessing steps for mixtral\_8x7b were:

\begin{itemize}
    \item Duplicate entries and null values were removed.
    \item Minimal text cleaning was applied, as the model can handle raw text effectively.
\end{itemize}

The model processes each document individually with the help of a prompt. The LLM analyzes the document's contents and identifies the core themes, generating five contextually relevant topics. This eliminates the need for dataset-wide analysis or clustering, focusing instead on each document's unique context. The following prompt was used to guide the LLM in extracting relevant topics from every document:

\begin{quote}
You are a topic modeling bot who will assign precise topics for the given input document. Given the document, assign 5 topics for the document. You must respond in the following manner: 'Topic number: Sentence of document that resulted in deriving the following topic'.
\end{quote}

\subsection{Proposed Methodology}

The proposed workflow starts by extracting the 'Threat' feature, ensuring that only unique entries are retained and null values are removed. Model-specific preprocessing techniques are then applied, where special characters, punctuation, and unnecessary noise are cleaned from the data to ensure smooth compatibility with the various topic modeling approaches.

Once preprocessing is completed, five topic modeling approaches—BERTopic with multiple configurations, CombinedTM, Top2Vec, Llama2 with BERTopic, and mixtral\_8x7b—are applied. Each model processes the data based on its working principles to extract latent topics or generate labeled clusters.

After the topics are extracted, the results from all models are evaluated using performance metrics like topic coherence, which measures how interpretable the topics are, and clustering quality, which assesses the compactness and separability of clusters. Visualizations, such as bar charts for top topic words, heatmaps to show topic similarity, and hierarchical cluster trees to reveal relationships between topics, are used to help analyze the outputs.

Lastly, a comparative analysis is conducted to identify the strengths and weaknesses of each approach. The goal of this workflow is to identify the best possible approach for categorizing software vulnerabilities.

\begin{figure}[htbp]
\centerline{\includegraphics[width=\linewidth]{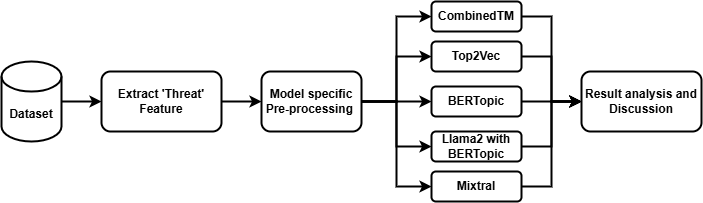}}
\caption{Proposed Workflow}
\label{fig:proposed_workflow}
\end{figure}

\section{Result Analysis and Discussion}

\subsection{Dataset Description and Visualization}

The dataset is collected from Cisco Labs and consists of 69,909 entries and 39 columns. This dataset provides an overview of software vulnerabilities, their vulnerability type and impact, and suggested remediation. It helps identify patterns and trends in vulnerabilities that could be used to enhance security measures and prioritize risks. This paper focuses on the \textit{‘Threat’} feature, which contains detailed textual descriptions of the vulnerabilities. This feature is the target input for topic modeling and clustering in the subsequent analysis.

\begin{figure}[htbp]
\centerline{\includegraphics[width=0.4\textwidth]{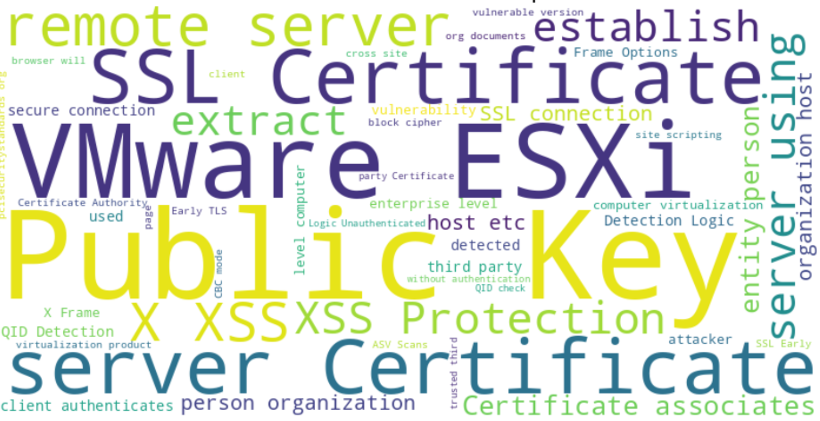}} 
\caption{Word Cloud of ‘Threat’ Descriptions}
\label{fig:wordcloud}
\end{figure}

Figure~\ref{fig:wordcloud} shows the word cloud generated from the \textit{‘Threat’} column, highlighting the top 50 most frequent terms. The visualizations highlight recurring terms such as "Public Key", "SSL Certificate", "VMware", and "ESXi," which indicate the prevalence of vulnerabilities related to server configurations, encryption protocols, and virtualization systems.

\subsection{Model specific results}
\subsubsection{BERTopic}
Several experiments were carried out using \textbf{BERTopic}, exploring various dimensionality reduction techniques, including \textbf{PCA} and \textbf{UMAP}. Additionally, we evaluated different clustering models, such as \textbf{HDBSCAN} and \textbf{KMeans}.

Hyperparameter tuning was carried out for the following components:

\begin{itemize}
    \item \textbf{Language Embeddings:} Improved the quality of textual representation by capturing semantic nuances, thereby impacting topic modeling performance.
    \item \textbf{Number of Topics:} Balanced granularity, with fewer topics grouping themes broadly and more topics providing specificity but risking redundancy.
    \item \textbf{Top Words:} Enhanced topic interpretability by selecting meaningful words to define topics.
    \item \textbf{Words Universe:} Controlled vocabulary scope, balancing diversity and noise.
\end{itemize}

This tuning optimized \textbf{model performance}, \textbf{interpretability}, and \textbf{topic coherence}.

\textbf{BERT Topic:} \\
The \textbf{UMAP + HDBSCAN} combination effectively identifies meaningful clusters, with hierarchical relationships providing additional insights into broader and more specific topic groupings.

\begin{figure}[h!]
    \centering
    \includegraphics[width=0.5\textwidth]{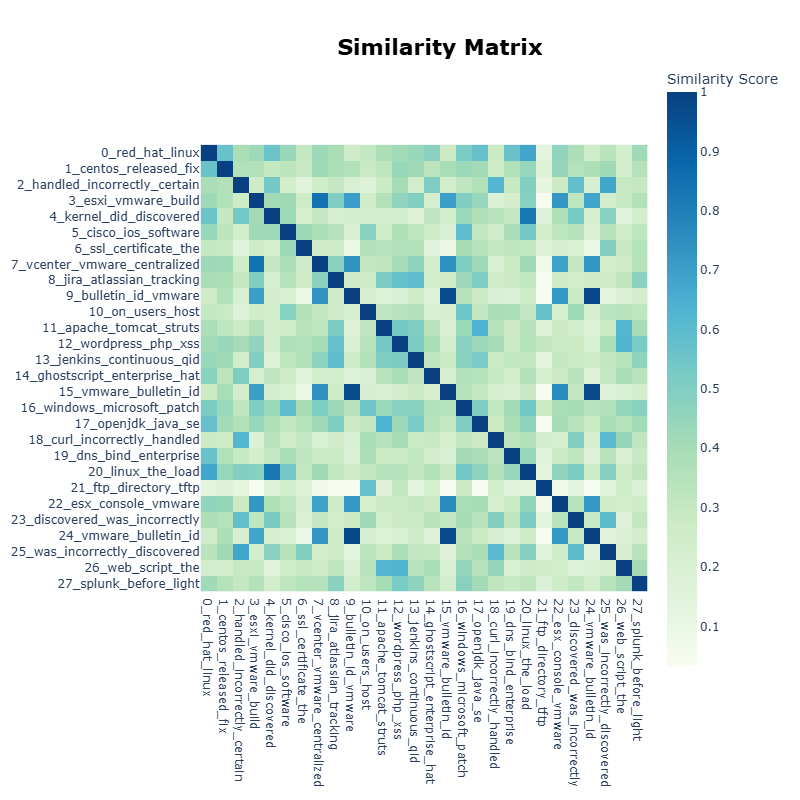} 
    \caption{Similarity Matrix of BertTopic using UMAP and HDBSCAN: Visualizing inter-topic relationships.}
    \label{fig:similarity_matrix}
\end{figure}

The similarity matrix (Figure \ref{fig:similarity_matrix}) provides an overview of the inter-topic relationships. The diagonal elements (similarity of a topic to itself) are understandably the highest, with scores close to 1. Off-diagonal elements indicate how closely related different topics are. Topics with higher similarity scores (darker shades) suggest shared or overlapping features in the underlying text data. 

For example:
\begin{itemize}
    \item Topics related to specific operating systems or software vulnerabilities may exhibit strong connections due to shared attributes in threat descriptions.
    \item Conversely, lighter regions highlight dissimilar topics, reflecting distinct vulnerability features or different contextual focuses within the dataset.
\end{itemize}

\begin{figure}[h!]
    \centering
    \includegraphics[width=0.5\textwidth]{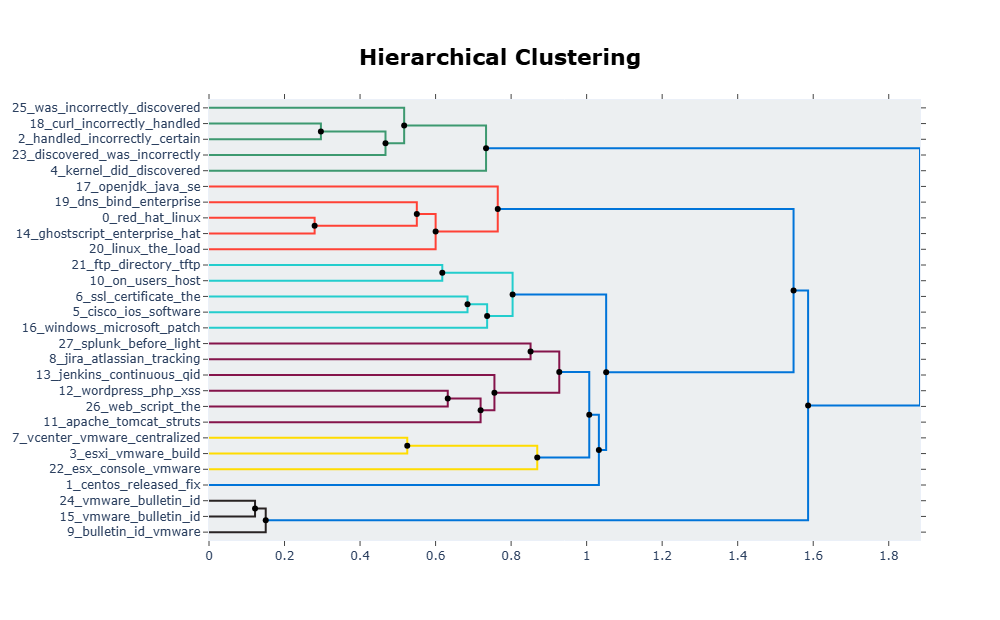} 
    \caption{Hierarchical Clustering: Visualizing topic relationships through a dendrogram.}
    \label{fig:hierarchical_clustering_UD}
\end{figure}

Hierarchical clustering (Figure \ref{fig:hierarchical_clustering_UD}) provides a clear representation of the relationships between topics, offering valuable insights into their grouping and hierarchical structure. Closely related topics, such as Topics 0, 1, and 2, are observed to form cohesive clusters. These clusters likely correspond to specific themes, such as vulnerabilities associated with Linux or CentOS systems, as suggested by the distribution of top words within each topic.

The height of the dendrogram branches in the hierarchical clustering plot signifies the degree of dissimilarity between clusters:
\begin{itemize}
    \item Topics with shorter branch connections merge at lower levels of the dendrogram, indicating closer relationships.
    \item Topics with longer branches reflect more distinct groupings.
\end{itemize}
This visualization aids in distinguishing between highly similar and more dissimilar topics.

The hierarchical arrangement of topics enables the identification of broader themes, such as operating system vulnerabilities, and their corresponding sub-themes, such as:
\begin{itemize}
    \item Linux kernel issues.
    \item SSL certificate weaknesses.
\end{itemize}
This structured approach provides a deeper understanding of the relationships between topics, highlighting both thematic overlaps and distinctions within the dataset.

The similarity heatmap and hierarchical clustering indicate strong grouping of topics related to similar types of vulnerabilities, such as operating system vulnerabilities (e.g., Linux, VMware) or specific threat types (e.g., SSL certificates).\\

\textbf{Bert Topic with Stacked Embeddings + UMAP:} Using the BERTopic model with stacked embeddings, we combine RoBERTa for contextual semantics and GloVe for word-level semantics, enhancing topic granularity and robustness. Dimensionality reduction is achieved using UMAP, which preserves the local structure of high-dimensional embeddings, while HDBSCAN identifies topic groupings without requiring a predefined number of clusters.

\begin{figure}[h!]
    \centering
    \includegraphics[width=0.5\textwidth]{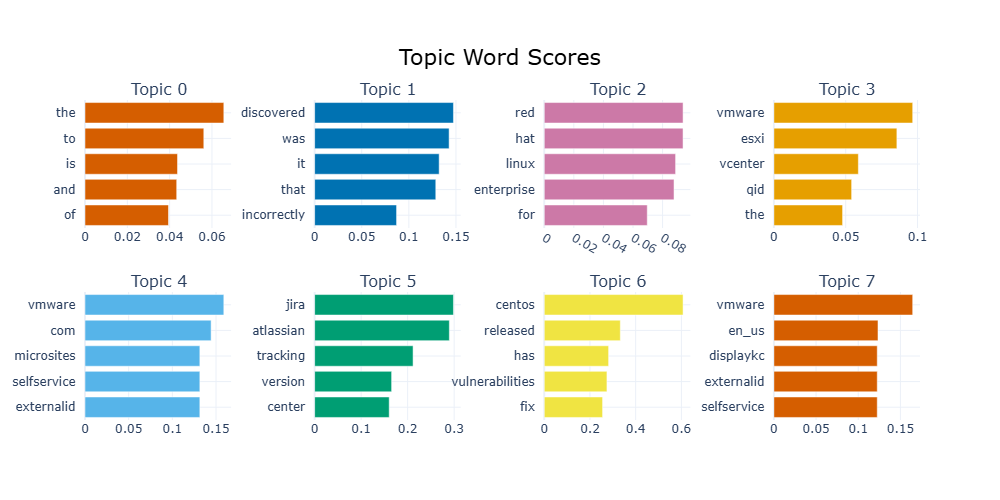} 
    \caption{Topic Word Score Bar Chart: Visualizing significant words for each topic.}
    \label{fig:topic_word_score}
\end{figure}

Referring to the Topic Word Score bar chart (Figure \ref{fig:topic_word_score}), the visualization highlights the most significant words for each topic, indicating their dominance in defining the respective topics. Below are the key observations:

\begin{itemize}
    \item \textbf{Topic -1:} Words like \textit{centos}, \textit{released}, and \textit{vulnerabilities} suggest a focus on CentOS-related security updates and fixes.
    \item \textbf{Topic 0:} Common words such as \textit{the}, \textit{to}, and \textit{is} indicate general or introductory text without a specific focus.
    \item \textbf{Topic 1:} Keywords like \textit{discovered}, \textit{handled}, and \textit{kernel} emphasize the identification and management of vulnerabilities in system kernels.
    \item \textbf{Topic 2:} Terms such as \textit{red}, \textit{hat}, \textit{linux}, and \textit{enterprise} demonstrate a strong focus on Red Hat Enterprise Linux vulnerabilities.
    \item \textbf{Topic 3:} Words like \textit{vmware}, \textit{esxi}, and \textit{vcenter} highlight vulnerabilities associated with VMware technologies.
    \item \textbf{Topics 4, 7, and 9:} Keywords such as \textit{vmware}, \textit{microsites}, \textit{selfservice}, and \textit{externalid} point to VMware knowledge base articles and support microsites.
    \item \textbf{Topic 5:} Terms like \textit{jira}, \textit{atlassian}, and \textit{tracking} indicate a focus on Atlassian JIRA versions and issue-tracking systems.
    \item \textbf{Topics 6, 8, and 10:} These topics consistently highlight CentOS updates and fixes, with representative words such as \textit{security}, \textit{update}, \textit{released}, and \textit{fix}.
\end{itemize}

These topic representations align with key software and vulnerability clusters, facilitating a better understanding of the thematic areas identified by the model.

\begin{figure}[h!]
    \centering
    \includegraphics[width=0.5\textwidth]{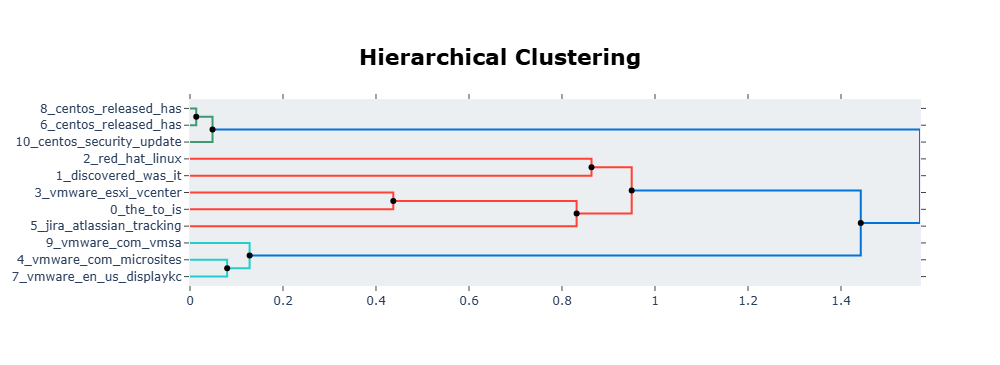} 
    \caption{Hierarchical Clustering Dendrogram: Visualizing topic relationships and their thematic hierarchies.}
    \label{fig:hierarchical_clustering_e}
\end{figure}

The hierarchical clustering (Figure \ref{fig:hierarchical_clustering_e}) of this approach reveals the relationships between topics based on similarity:

\begin{itemize}
    \item \textbf{Clusters:} Topics 0, 1, 2, and 3 form a closely related group, likely sharing overlapping themes, such as Linux or VMware-related vulnerabilities.
    \item \textbf{Dissimilarity:} Topics such as 6 and 7 merge at higher levels, reflecting distinct thematic focuses, possibly on software updates or other unrelated contexts.
    \item \textbf{Branch Height:} Shorter branches (e.g., Topics 2 and 3) signify closely related topics, while longer branches (e.g., Topic 0 vs. Topic 7) highlight more distinct themes.
\end{itemize}

This clustering highlights thematic hierarchies, enabling a broader understanding of topic relationships.

\begin{figure}[h!]
    \centering
    \includegraphics[width=0.5\textwidth]{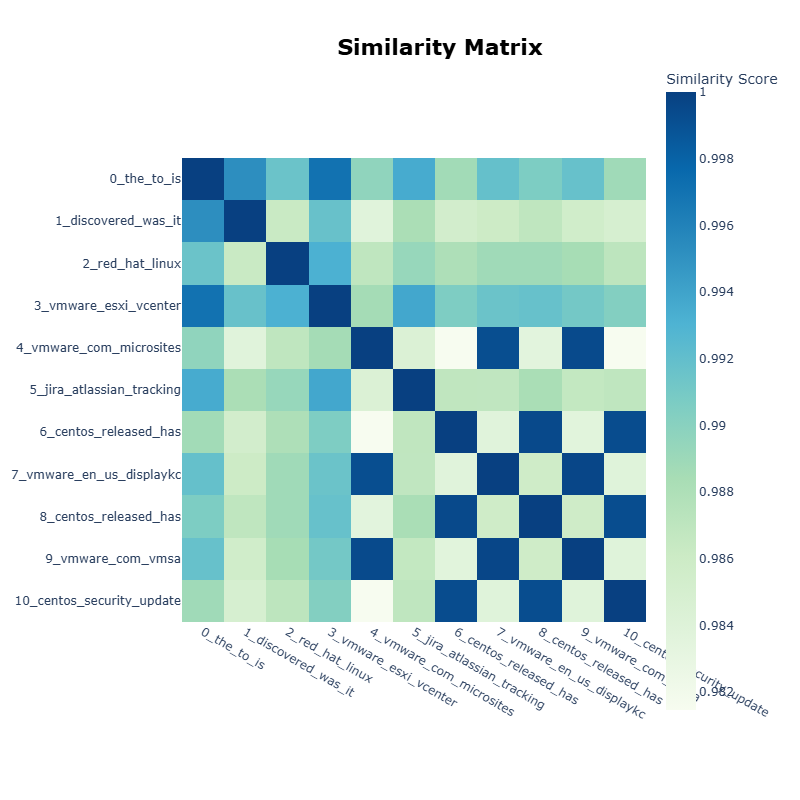} 
    \caption{Heatmap: Visualizing thematic overlaps and distinctions between identified topics.}
    \label{fig:heatmap_e}
\end{figure}

The heatmap (Figure \ref{fig:heatmap_e}) provides insights into the relationships between topics, revealing thematic overlaps and distinctions. For instance, Topics 2 and 3 exhibit strong similarity, likely due to shared references to VMware or Linux systems, while other topics, such as Topic 0 and Topic 7, appear more distinct, reflecting divergent themes. This visualization validates the clustering results, confirming the coherence of closely related topics and the separation of those with minimal overlap. The heatmap effectively supports the hierarchical clustering and topic word scores, highlighting the consistency and robustness of the identified thematic groupings.

\subsubsection{Mistral AI LLM for topic assignment}
Utilising Mistral LLM for topic modeling was a straightforward approach, the model successfully identified the main topics within each document, demonstrating its ability to extract meaningful insights from complex text. This straightforward approach leverages the strengths of large language models to streamline the process of topic modeling, making it a valuable tool for various text analysis tasks. 

\subsubsection{Combined TM}
\begin{figure}[h!]
    \centering
    \includegraphics[width=0.5\textwidth]{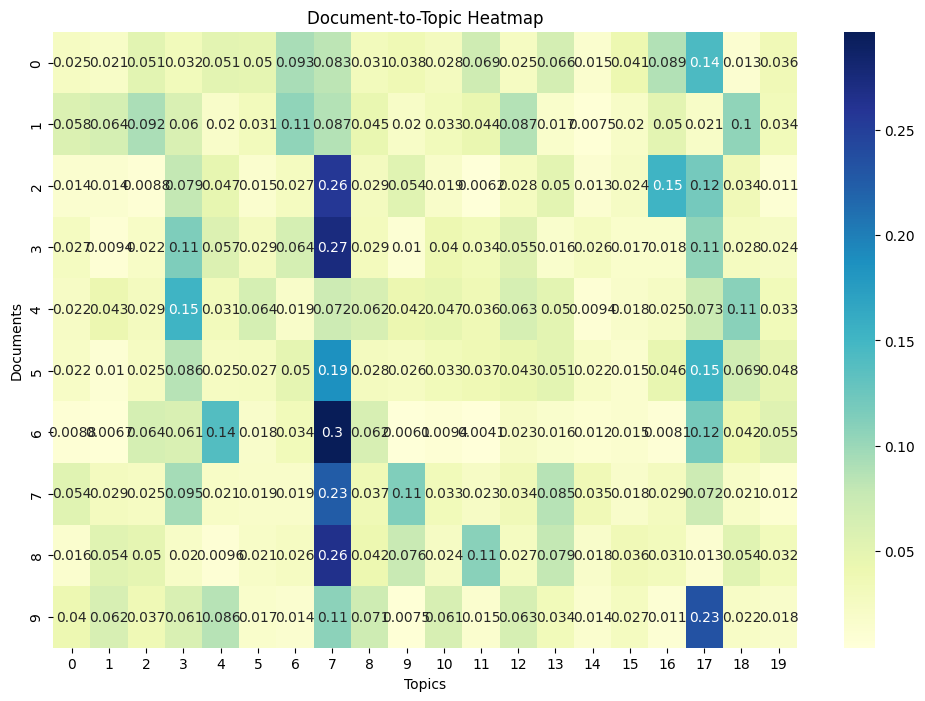}
    \caption{Heatmap: Visualizing thematic overlaps and distinctions between identified topics.}
    \label{fig:Com_heat}
\end{figure}

\begin{figure}[h!]
    \centering
    \includegraphics[width=0.5\textwidth]{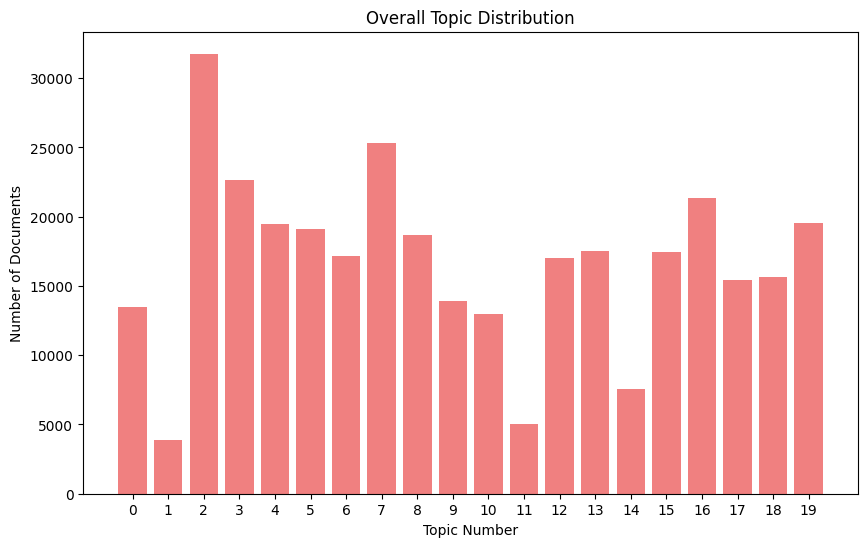}
    \caption{Heatmap: Visualizing thematic overlaps and distinctions between identified topics.}
    \label{fig:Com_dist}
\end{figure}

The results of the Combined Topic Model reveal notable improvements in coherence and interpretability of topics through the integration of BoW representations into the SBERT contextualized embeddings. The document-to-topic heatmap in Figure \ref{fig:Com_heat} reveals clear differentiation between documents and their correlated topics, showing that the model effectively captures unique thematic clusters. The overall topic distribution in Figure \ref{fig:Com_dist} graph reveals that while some topics (e.g., Topic 2 and Topic 7) are more prevalent across the dataset, others maintain a balanced presence, indicating diversity in topic coverage.

Overall, the CombinedTM approach efficiently enhances topic coherence and clarity, leveraging contextual embeddings to overcome the limitations of traditional methods of topic modeling. This synergy between BoW and contextual representations ensures robust thematic analysis, allowing this model to be applied to the analysis of highly diversified as well as complex text corpora.
\subsubsection{Top2Vec}
\begin{figure}[h!]
    \centering
    \includegraphics[width=0.5\textwidth]{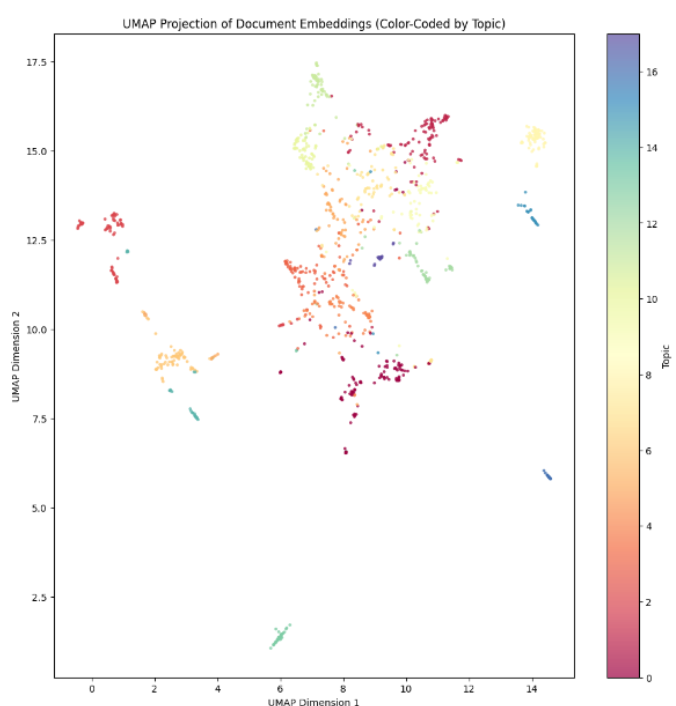}
    \caption{UMAP projection of Document Embeddings}
    \label{fig:top2vec}
\end{figure}

Figure \ref{fig:top2vec}, the UMAP projection colors high-dimensional document embeddings by their topic assignments. Well-separated clusters, such as Topics 2, 6, and 14, are internally coherent, while proximal clusters, such as Topics 5, 7, and 10 reflect partial thematic overlap. Isolated points likely represent outliers or very small topics, such as Topic 1, which only contained one document.
Larger topics, in the sense of Topic 6 and Topic 14, are dominant, although smaller topics, such as Topics 1 and 15, might be niche issues within more general topics. Truly generic topics, like the ones on 4, 8, and 12 featuring terms such as "vulnerabilities," should be further honed or combined to better result in a granular composition. Technical topics, going to Topic 12 focusing on SSL/TLS vulnerabilities or even Topic 17, representing DNS and server issues, are logical and representative.

Smaller or overlapping topics could be analyzed and combined to enhance coherence. Inclusion of domain-specific stopwords during preprocessing might reduce redundancy in "vulnerability"-centric topics, while document-level analysis may validate thematic consistency. The UMAP visualization aligns well with the topic data; in the data, dominant clusters and scattered points for rare or noisy topics indicate further refinement opportunities.
\subsubsection{Llama2 with BERTopic}
The integration of Llama2 with BERTopic successfully processed the dataset and identified three distinct topics, summarized in the table below:

\begin{table}[htbp]
\caption{Topics Generated for Llama2 with BERTopic}
\begin{center}
\renewcommand{\arraystretch}{1.2} 
\begin{tabular}{|c|c|l|}
\hline
\textbf{Topic ID} & \textbf{Count} & \textbf{Name} \\ 
\hline
-1 & 287 & -1\_vmware\_esxi\_com\_http \\ 
\hline
0 & 572 & 0\_red\_hat\_linux\_enterprise \\ 
\hline
1 & 267 & 1\_discovered\_incorrectly\_handled\_certain \\ 
\hline
\end{tabular}
\label{tab:llama2_bertopic_topics}
\end{center}
\end{table}

Topic -1, consisting of 287 documents, primarily represents outliers containing VMware-related keywords and generic HTTP references, which are less cohesive compared to the other clusters. Topic 0 is the largest cluster with 572 documents, demonstrating a strong focus on Red Hat Linux Enterprise, capturing dominant themes and high-frequency topics within the dataset. Topic 1, with 267 documents, represents specific and niche technical issues, such as kernel driver errors and incorrect handling of certain operations, providing valuable insights into critical technical challenges. The overall performance of Llama2 with BERTopic highlights its robustness in efficiently clustering diverse datasets into meaningful groups. It not only identifies central themes but also isolates less relevant or noisy data into distinct outlier clusters. This capability makes the integration particularly suitable for analyzing technical datasets and deriving actionable insights. The table provides a clear overview of the distribution and significance of the topics extracted, ensuring transparency and interpretability of the results.

\section{Conclusion and Future Work}
This study demonstrates the potential of advanced topic modeling techniques and large language models (LLMs) in addressing the challenges posed by the growing complexity and volume of software vulnerabilities [17]. By focusing on the 'Threat' feature of a real-world dataset, models such as BERTopic, Top2Vec, CombinedTM, Llama2 with BERTopic, and Mixtral were employed to extract meaningful patterns and generate interpretable clusters. The integration of dimensionality reduction techniques like UMAP and PCA, along with clustering methods such as HDBSCAN and DBSCAN, significantly enhanced the granularity and coherence of the identified topics. The comparative analysis of these models revealed their strengths in uncovering latent themes and prioritizing vulnerabilities based on contextual relevance. The proposed approaches not only streamline the process of analyzing large datasets but also provide actionable insights, enabling cybersecurity professionals to better allocate resources and mitigate risks.

This research contributes to the development of scalable and automated solutions for software vulnerability management, making it easier to identify and address critical threats. Future work could explore the incorporation of real-time processing capabilities, the adaptation of these techniques to multilingual datasets, and the integration of predictive modeling for enhanced threat forecasting. By bridging theoretical advancements in topic modeling with practical cybersecurity applications, this study paves the way for more efficient and robust vulnerability detection systems.


\begin{thebibliography}{00}
\bibitem{b1} Reuter, Arik, et al. "GPTopic: Dynamic and Interactive Topic Representations." arXiv preprint arXiv:2403.03628 (2024).
\bibitem{b2} Rijcken, Emil, et al. "Towards interpreting topic models with ChatGPT." The 20th World Congress of the International Fuzzy Systems Association. 2023.
\bibitem{b3} Petukhova, Alina, Joao P. Matos-Carvalho, and Nuno Fachada. "Text clustering with LLM embeddings." arXiv preprint arXiv:2403.15112 (2024).
\bibitem{b4} Frei, Stefan, et al. "Large-scale vulnerability analysis." Proceedings of the 2006 SIGCOMM workshop on Large-scale attack defense. 2006.
\bibitem{b5} Weider, D. Yu, Dhanya Aravind, and Passarawarin Supthaweesuk. "Software Vulnerability Analysis for Web Services Software Systems." iscc. 2006.
\bibitem{b6} Akash, Pritom Saha, and Kevin Chen-Chuan Chang. "Enhancing Short-Text Topic Modeling with LLM-Driven Context Expansion and Prefix-Tuned VAEs." arXiv preprint arXiv:2410.03071 (2024).
\bibitem{b7} Schneider, Johannes. "Topic Modeling with Fine-tuning LLMs and Bag of Sentences." arXiv preprint arXiv:2408.03099 (2024).
\bibitem{b8} Mu, Yida, et al. "Large Language Models Offer an Alternative to the Traditional Approach of Topic Modelling." arXiv preprint arXiv:2403.16248 (2024).
\bibitem{b9} Gana, Bady, et al. "Leveraging LLMs for Efficient Topic Reviews." Applied Sciences 14.17 (2024): 7675.
\bibitem{b10} Yang, Xiaohao, et al. "LLM Reading Tea Leaves: Automatically Evaluating Topic Models with Large Language Models." arXiv preprint arXiv:2406.09008 (2024).
\bibitem{b11} Sandilya, Harshit, et al. "Generating topic-agnostic conversations With LLMs." IEEE access (2024).
\bibitem{b12} Zeng, Peng, et al. "Software vulnerability analysis and discovery using deep learning techniques: A survey." IEEE Access 8 (2020): 197158-197172.
\bibitem{b13} Williams, Mark A., et al. "A vulnerability analysis and prediction framework." Computers \& Security 92 (2020): 101751.
\bibitem{b14} Niranjan D K, N Rakesh, “Real Time Analysis of Air Pollution Prediction using IoT”, in the 2nd International Conference on Inventive Research in Computing Application [ICIRCA 2020], July-2020.
\bibitem{b15} Niranjan D K, N Rakesh, “Smart Surveillance System by Face Recognition and Tracking using Machine Learning Techniques”, in the 4th International Conference on Computational Vision and Bio Inspired Computing [ICCVBIC 2020], November-2020.
\bibitem{b16} Sainadh. K.V., Satwik. K., Ashrith. V., Niranjan. D.K. (2023), “A Real-Time Human Computer Interaction Using Hand Gestures in OpenCV”, in the Information and Communication Technology for Intelligent Systems. ICTIS 2023. Lecture Notes in Networks and Systems, vol 720. Springer, Singa
\bibitem{b17} N. .D.K. and Rakesh, N., “Early Building Collapse Detection using IoT”, in International Conference on Inventive Research in Computing Applications [ICIRCA 2020] organized by RVS College of Engineering and Technology, Coimbatore, India, 2020.
\bibitem{b18} S. M, R. V. Savant, S. Seshadri, N. Narmada and P. B. Pati, "Unveiling Hidden Patterns: Clustering Algorithms on C Code embedding," 2024 IEEE 9th International Conference for Convergence in Technology (I2CT), Pune, India, 2024, pp. 1-7, doi: 10.1109/I2CT61223.2024.10543306.
\bibitem{b19} Rao, S.S., Mishra, S., Akhilesh, S., Balakrishnan, R.M. and Pati, P.B., 2024, June. Automatic Assessment of Quadratic Equation Solutions Using MathBERT and RoBERTa Embeddings. In 2024 15th International Conference on Computing Communication and Networking Technologies (ICCCNT) (pp. 1-7). IEEE.
\bibitem{b20}Anirudh, S., Nishant, P.R., Baitha, S. and Kumar, K.D., 2024. An Ensemble Classification Model for Phishing Mail Detection. Procedia Computer Science, 233, pp.970-978.

\end{thebibliography}
\end{document}